\documentclass[aps,prd,twocolumn,superscriptaddress]{revtex4}

\usepackage{epsfig,epsf}
\usepackage{amsmath}
\usepackage{amsthm}
\usepackage{amsfonts}
\usepackage{amssymb}
\usepackage{dsfont}
\usepackage{multirow}
\usepackage{appendix}
\usepackage{slashed}
\usepackage[active]{srcltx}
\usepackage{psfrag}
\usepackage{graphicx}
\usepackage{multirow}
\usepackage{subfigure}
\input epsf

\usepackage{subfigure}
\begin{document}   

\title{ S-Wave Single Heavy Baryons with Spin-3/2 at Finite Temperature}
\date{\today}
\author{K. Azizi$^{1,2}$, A. T{\"u}rkan$^{3}$  \\
\textit{$^{1}$Department of Physics, University of Tehran, North Karegar Avenue, Tehran 14395-547, Iran\\
$^{2}$ Department~of~Physics,~Do\v{g}u\c{s} University, Ac{\i}badem-Kad{\i}k\"oy,  34722 Istanbul, Turkey\\
$^{3}$ \"Ozye\v{g}in University, Department~of Natural~and~Mathematical~Sciences, \c{C}ekmek\"{o}y, 34794 Istanbul, Turkey}}

\begin{abstract}
The thermal behavior of the  spectroscopic parameters of the S-wave  single heavy baryons  $ \Sigma_{Q}^{*}, \Xi_{Q}^{*}$ and $ \Omega_{Q}^{*} $  with spin-3/2 are investigated in QCD at finite temperature. We analyze the variations of the mass and residue of these baryons taking into consideration the  contributions of QCD thermal condensates up to dimension eight in Wilson expansion. At finite  temperature, due to the breakdown of the Lorentz invariance by the 
choice of reference frame and presence of an extra $O(3)$ symmetry, some new four-dimensional operators come out in the form of the fermionic and gluonic parts of the energy momentum tensor that are taken into account in the calculations. Our analyses show that at lower temperatures, the parameters of baryons under consideration are not affected by the medium. These parameters, however, show rapid variations with respect to temperature at higher temperatures near to a pseudo-critical temperature, after which the baryons are melted. The results of the masses and residues at $ T\rightarrow 0 $ limit are  compatible with the available  experimental data and predictions of  other theoretical studies.
\end{abstract}

\maketitle



%

\section{Introduction}
\begin{figure}[ht]
	\begin{center}
		\includegraphics[width=5.5cm]{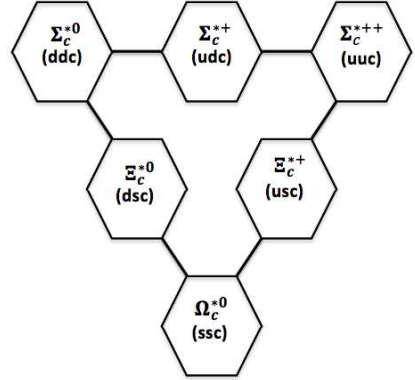}		
\end{center}
	\caption{The sextet representation of single charmed  baryons with total spin-$ 3/2 $. The same picture is valid for bottom baryons with the replacement $ c\rightarrow b $.} \label{fig1}
\end{figure}

With the rising number of experimental data on charmed and bottom baryons, the interest in the investigation of heavy baryons has increased, considerably. Before giving the details of the experimental studies on heavy baryons, it would be useful to give some theoretical information. The Quark Model is one of the most successful tools to classify the mesons and baryons. The traditional single heavy baryons ($ Qqq $) consist of one heavy ($ Q=b$ or  $c $) and two light quarks ($ q=u, d$ or  $ s $). The mass of heavy quark is very large compared to the light quark masses and the light degrees of freedom form a diquark $ qq $, which orbits the nearly static heavy $ Q $  quark. Therefore, infinitely heavy mass limit ($ m_{Q}\rightarrow\infty $) for the heavy quark is utilized to classify the single heavy baryons \cite{Isgur,Georgi}. In this case, for the two light quarks, the total flavor-spin wave function has to be symmetric because  their color wave function is antisymmetric. Hence there are two different representations for the S-wave heavy baryons ($\textbf{3} \otimes \textbf{3}=\overline{\textbf{3}} \oplus \textbf{6}$): antisymmetric $  \overline{\textbf{3}} $ or symmetric $ \textbf{6}$. The antitriplet ($ \overline{\textbf{3}} $) of baryons contain only spin-$ 1/2 $ states while the sextet ($ \textbf{6} $) of baryons contain both spin-$ 1/2 $ and spin-$ 3/2 $ states.
In this study, we investigate the thermal properties of the single heavy bottom/charmed spin-$ 3/2 $ sextet states:  The members for charmed baryons are shown in Figure \ref{fig1}.

Experimentally, the $ \frac{1}{2}^{+} $ antitriplet ($ \Lambda^{+}_{c}, \Xi^{+}_{c}, \Xi^{0}_{c} $) states, the $ \frac{1}{2}^{+} $ sextet ($ \Omega_{c}, \Sigma_{c}, \Xi^{\prime}_{c} $) baryons and the $ \frac{3}{2}^{+} $ sextet ($ \Omega^{*}_{c}, \Sigma^{*}_{c}, \Xi^{*}_{c} $) resonances have been observed in the charmed sector while the only $  \Lambda_{b}, \Sigma_{b}^{(*)}, \Xi_{b}^{(*)}$ and $ \Omega_{b} $ have been discovered in the bottom picture \cite{Nakamura}. Some history of discoveries are in order:  In 2006 the CDF collabration reported observation of  $  \Lambda_{b} $ \cite{CDF} and $ \Omega^{*}_{c} $ discovered by the Babar  collaboration \cite{Babar}. The CDF collabration reported the first observation of $ \Sigma_{b} $ and $ \Sigma^{*}_{b} $ baryons later \cite{CDF2}. The D0 collabration declared the observation of $ \Xi_{b} $ \cite{D0} and it was confirmed by  CDF in a short time \cite{CDF3}. The observation of ground and excited states of $ \Xi_{c} $  were proclaimed by Belle and BABAR collabrations  \cite{Chistov, Aubert}. $ \Xi^{*}_{c} $ observed by Belle in 2008 \cite{Lesiak} and  discovery of $ \Xi^{*}_{b} $ was reported by CMS and LHCb collaborations \cite{CMS, LHCb}.

 On the other hand, various theoretical studies in vacuum have been utilized to investigate the spectroscopic parameters of single heavy baryons. In 1982 Shuryak primarily calculated the heavy baryon masses via QCD sum rule \cite{Shuryak}. In Ref. \cite{Capstick} Capstick and Isgur examined the heavy baryon systems in a quark potential model. Bagan  et al. investigated the heavy baryons by taking  into account the separation of negative and positive parity contributions \cite{Bagan}. Grozin and Yakovlev evaluated the masses of $\Lambda_{Q} $ and $ \Sigma^{(*)}_{Q} $ using the  heavy quark effective theory (HQET) \cite{Grozin}. Charmed baryons were investigated in Chiral perturbation theory by Savage and also results were extended for b-baryons in the same study \cite{Savage}. Roncaglia et  al. in Ref \cite{Roncaglia} estimated the heavy baryon masses with one/two heavy quark/quarks in the framework of  Feynman-Hellman theorem. In Ref.  \cite{Jenkins} Jenkins studied the masses of heavy baryons in the $ 1/m_{Q} $ and $ 1/N_{c} $ expansions. The $ 1/m $ corrections to heavy baryon masses were calculated by Dai et al. in the framework of the HQET \cite{Dai}. QCD sum rule for heavy baryons at leading order in $ 1/m_{Q} $ and at next to the leading order in $\alpha_{s}$ were evaluated by Groote et al. in Ref. \cite{Groote}. Wang et  al. improved the analysis for the $\Lambda_{Q} $ and $ \Sigma_{Q} $ baryon masses to order $ \Lambda_{QCD}/m_{Q}$ from QCD sum rule \cite{Wang}. Mathur et al. predicted the mass spectrum of charmed and bottom baryons from Lattice QCD \cite{Mathur}. Wang and Huang in Ref. \cite{Wang1} studied the mass, coupling constant, and Isgur-Wise function for ground-state heavy baryons within the framework of HQET by taking into account both the two and three-point correlation functions. Ebert et  al. computed heavy baryon masses in the heavy-quark light-diquark approximation in the framework of constituent quark model \cite{Ebert}. Garcilazo et  al. solved exactly the three quark problem via Faddeev method in momentum space \cite{Garcilazo}. In Ref. \cite{Zhang} Zang and Huang calculated the charm and bottom baryon masses up to operator dimension six in operator product expansion (OPE) by the help of the QCD sum rule approach. The mass and residue of $ \Omega^{*}_{c} $ and $ \Omega^{*}_{b} $ with spin parity $ 3/2^{+} $ were studied by Wang via QCD sum rule \cite{Wang2}. A quark model was applied to the spectrum of baryons containing one heavy baryon by Roberts and Pervin \cite{Roberts}. Bottom baryon spectra were investigated using Faddeev method in momentum space by Valcarce et al. \cite{Valcarce}. Liu et  al. performed a systematic study of the masses of bottom baryons up to $ 1/m_{Q} $ in HQET  \cite{Liu}. Groote et al. computed the NLO perturbative corrections for the static properties of heavy baryons \cite{Groote1}. In Ref. \cite{Zhang1} the mass of $\Lambda_{Q} $ and $ \Sigma^{(*)}_{Q} $ baryons were calculated by Zhang and Huang via QCD sum rule taking into account operators up to dimension six. Using the coupled channel formalism, Gerasyuta and Matskevich calculated the S-wave bottom baryons masses \cite{Gerasyuta}. In Ref. \cite{Karliner} Karliner et  al. investigated the b-baryons in the quark model. In two-point and light cone QCD sum rule methods Aliev et al. studied the mass and magnetic moments of single heavy baryons with spin-$ 3/2 $ \cite{Aliev}. Lewis and Shyn predicted the bottom baryon masses based on a $ 2+1$ flavor dynamical lattice QCD simulation \cite{Lewis}. The spin-$ 3/2^{+} $ heavy and doubly heavy baryon states \cite{WangZG}  were investigated by subtracting the contributions from the corresponding negative parity by Z. G. Wang. The mass spectra of heavy baryons were studied by the help of the motivated relativistic quark model by Ebert \cite{Ebert1}. Kim et al.  investigated the single heavy baryon mass based on the self-consistent Chiral quark soliton model in Ref. \cite{Kim}. Finally,  Azizi and Er studied the in-medium properties of spin-$ 3/2 $ heavy baryons in nuclear matter using QCD sum rule in a dense medium \cite{Azizi}.

Theoretical investigations of spectroscopic parameters of the single heavy baryons at finite temperature will help us better understand and analyze the results of  heavy-ion collision experiments and gain valuable information  on the internal structures of these baryons, behavior of these  baryons near to a pseudo-critical temperature, possible phase transition/ crossover \cite{Aoki,MCheng} to/with quark gluon plasma (QGP) (adopted as a new phase of matter ) as well as the perturbative and nonperturbative dynamics of QCD. At extreme temperatures, two different possibilities can be considered: crossover and phase transition. Many Lattice calculations predict that crossover  occurs at $ T_{pc}\approx155 MeV$ \cite{Bhattacharya,Bazavov2}. For the QGP phase transition, we need greater temperature values and there is no unique temperature to the phase transition of QGP. At short distances, to describe the strong interaction QCD is a suitable theory. However, the calculations of hadronic parameters including nonperturbative effects (occur in low energy scale)  usually need some nonperturbative  phenomenological models. Many phenomenological models are available in the literature: QCD sum rule is one of the powerful ones among them. This method  firstly suggested by Shifman, Vainshtein and Zakharov to investigate the vacuum properties of mesons \cite{Shifman} and then Ioffe \cite{Ioffe} applied this method for baryons. The thermal version of the QCD sum rule was extended by Bochkarev and Shaposhnikov \cite{Bochkarev}. In addition to the vacuum expectation values of quark and gluon condensates, their thermal forms and some new operators appear in the thermal version. 

In this study, we investigate the temperature effects on the spectroscopic parameters of the ground state sextet baryons including single heavy quark and with spin-$ 3/2 $ via thermal QCD sum rule method. Taking into account the additional operators coming from OPE due to breaking of the Lorentz invariance by the choice of the thermal rest frame, condensates up to dimension eight are considered.
The article is arranged in the following form. In Sec. II, the in-medium sum rules  for the mass and residues of the $ \Sigma_{Q}^{*}, \Xi_{Q}^{*}$ and $ \Lambda_{Q}^{*} $  single heavy baryons are obtained. In Sec. III the numerical analysis for the spectroscopic parameters under consideration is performed. The last section includes the summary and our concluding remarks.


%

\section{Calculations }

\label{sec:DA} %

In this section, QCD sum rules for the spectroscopic parameters of the spin-3/2 $\Sigma^{*}_Q$, $\Xi^{*}_Q$ and $\Omega^{*}_Q$ baryons  are obtained at finite temperature. To this end, we start with the following two-point thermal correlation function:
\begin{equation}\label{CorrFunc}
\Pi_{\mu \nu }(q,T)=i\int d^{4}x~e^{iq\cdot x}\langle
\Psi|\mathcal{T}\{J_{\mu}(x) \bar{J}_{\nu}(0)\}|\Psi\rangle,
\end{equation}
where $q$ is the four-momentum of the chosen baryon, $\Psi $ is the ground state of the hot medium, $\mathcal{T}$ 
denotes the time-ordering operator and $J_{\mu}(x)$ is the interpolating current of the single heavy baryon, $ B_{SH} $. 
%
%

As the standard procedures of the QCD sum rule, the correlation function given above can be calculated at different contexts. At large distances, it is evaluated in terms of the hadronic parameters such as the mass and  residue of hadron. We call it the physical or hadronic representation of the correlator. The same correlator can be expressed in terms of the quark, gluon and mixed condensates by the help of the OPE at $ q^{2}<< 0 $ region. The computations in this way contain short distance effects. This representation, is generally called the OPE or QCD side of the correlation function. Finally, we match the two windows and compare the coefficients of the  same Lorentz structures from both sides. To remove the unwanted contributions coming from the higher states and continuum, Borel transformation as well as continuum subtraction, supplied by the quark-hadron duality assumption at finite temperature, are performed.  These procedures bring some auxiliary parameters, which we fix them before making any numerical estimations on the physical quantities.

To obtain the physical side of the correlator, a complete set of intermediate state with the same quantum numbers and quark content as the chosen current is inserted between the interpolating currents in correlation function. This is followed by the integral over four-$x$, which leads to
\begin{eqnarray}\label{piphys}
&&\Pi_{\mu\nu}^{Phys}(q,T)\notag \\&=&-\frac{{\langle}\Psi|J_{\mu}(0)|B_{SH}(q,s){\rangle}
{\langle}B_{SH}(q,s)|J^{\dag}_{\nu}(0)|\Psi{\rangle}}{q^{2}-m_{B_{SH}}^{2}(T)} \notag \\
&+&\mbox{contribution~of~higher~states~and~continuum},\notag \\
\end{eqnarray}
where $m_{B_{SH}}(T)$ is the temperature-dependent mass of the ground state of $ B_{SH} $. The matrix element  ${\langle}\Psi|J_{\mu}(0)|B_{SH}(q,s){\rangle}$ is defined in terms of the temperature dependent residue, $ \lambda_{B_{SH}}(T) $, as 
\begin{eqnarray}\label{matrixelement}
{\langle}\Psi|J_{\mu}(0)|B_{SH}(q,s){\rangle}&=&\lambda_{B_{SH}}(T)u_{\mu}(q,s) ,
\end{eqnarray}
where $u_{\mu}(q,s)$ is the Rarita-Schwinger spinor. The final form of the physical side can be obtained by inserting Eq. (\ref{matrixelement}) into Eq. (\ref{piphys}) and summing over the spins of the $ B_{SH} $. The summation over Rarita-Schwinger spinors is performed using
\begin{eqnarray}\label{Rarita}
&&\sum_{s}u_{\mu}(q,s)\bar u_{\nu}(q,s)\notag \\&=&-\Big( \!\not\!{q}+m_{B_{SH}}\Big)\Big[g_{\mu\nu} - \frac{1}{3}\gamma_{\mu}\gamma_{\nu}\notag \\
&-&\frac{2\,q_{\mu}q_{\nu}}{3m^{2}_{B_{SH}}} + \frac{q_{\mu}\gamma_{\nu}-q_{\nu}\gamma_{\mu}}{3m_{B_{SH}}}\Big].
\end{eqnarray}
By using the above behest, we recast the physical side as
\begin{eqnarray}
\label{}
\Pi^{Phys}_{\mu\nu}(q,T)&=&\frac{\lambda^{2}_{B_{SH}}(T)(\!\not\!{q}+m_{B_{SH}})}{q^{2}-m^{2}_{B_{SH}}} \Big[g_{\mu\nu} - \frac{1}{3}\gamma_{\mu}\gamma_{\nu}\notag \\
&-&\frac{2\,q_{\mu}q_{\nu}}{3m^{2}_{B_{SH}}} + \frac{q_{\mu}\gamma_{\nu}-q_{\nu}\gamma_{\mu}}{3m_{B_{SH}}}\Big]+ ...,
\end{eqnarray}
where $ \lambda^{2}_{B_{SH}}(T)=\lambda_{B_{SH}}(T)\bar{\lambda}_{B_{SH}}(T) $. It should also be specified that the interpolating current $ J_{\mu}(x) $ couples to both the spin-1/2 and spin-3/2 states. In this study, we only consider the contribution of spin-3/2 heavy baryons and we need to comb out the pollution of spin-1/2 state. These unwanted contributions can be eliminated in two different ways: 1) For spin-3/2 state, it should be introduced a projection operator which destroys the spin-1/2 contributions, 2)By a specific ordering of the Dirac matrices and remove the terms corresponding to the spin-1/2 particles (for more details  see for instance \cite{Aliev1}).The contribution of the spin-1/2 states can be traced using 
\begin{equation}
\label{spin12}
\langle\Psi|J_{\mu}(0)|\frac{1}{2}(q)\rangle=\Big[\kappa_1 q_{\mu}+\kappa_2 \gamma_{\mu}\Big] u(q),
\end{equation}
where $\kappa_1$ and $\kappa_2$ are some constants. By applying the condition $J_{\mu}\gamma^{\mu}=0$ (for more details see \cite{Savvidy}), we get $\kappa_1$ in terms of $\kappa_2$. Hence,
\begin{eqnarray}\label{intcur2}
{\langle}\Psi|J_{\mu}(0)|\frac{1}{2}(q){\rangle}&=&\kappa_2\Big(\gamma_{\mu}-\frac{4}{m_{\frac{1}{2}}} q_{\mu}\Big)u(q).
\end{eqnarray}
As is seen from Eq. (\ref{intcur2}), the pollution coming from spin-$ 1/2 $ resonances are commensurate to either $q_{\mu}$ or $\gamma_{\mu}$. To remove these contributions, the Dirac matrices are ordered as $\gamma_{\mu} \!\not\!{q}\gamma_{\nu}$ and terms proportional to $q_{\mu}$ or  $q_{\nu}$, also  those  beginning with $\gamma_{\mu}$ or ending with $\gamma_{\nu}$ are set to zero. Finally, the clean physical side of the correlator, in the Borel scheme, is  obtained as
\begin{eqnarray}\label{Physical}
\hat{B}\Pi_{\mu\nu}^{Phys}(q,T)&=&\lambda^{2}_{B_{SH}}(T)e^{-m_{B_{SH}}^{2}(T)/M^2}\!\not\!{q}g_{\mu\nu}\notag \\
&+& \lambda^{2}_{B_{SH}}(T)m_{B_{SH}}e^{-m_{B_{SH}}^{2}(T)/M^2}g_{\mu\nu}\notag \\
&+&...,
\end{eqnarray}
where $M^2$  is the Borel parameter  and dots denote the contributions of other structures as well as the higher states and continuum.

The next step is to calculate the OPE side of the correlation function. In  deep Euclidean region, the  correlation function is evaluated in terms of the quark and gluon degrees of freedom by the help of Wilson expansion. To achieve this goal, the basic point it to choose a  suitable interpolating current defining the particles under study. The  interpolating current for spin-$ 3/2 $ $ B_{SH} $  in a compact form can be written as \cite{Aliev2,Aliev3,Lee}
\begin{eqnarray}
J_{\mu}(x) &=&A~ \epsilon_{abc}\
\Bigg[\Big( q_1^{aT}(x)C\gamma _{\mu}q_2^{b}(x)\Big) Q^{c}(x) \notag \\
&+& \Big( q_2^{aT}(x)C\gamma _{\mu}Q^{b}(x)\Big) q_1^{c}(x) \notag \\
&+& \Big( Q^{aT}(x)C\gamma _{\mu}q_1^{b}(x)\Big) q_2^{c}(x) \Bigg],
\label{current}
\end{eqnarray}
where $A$ is the normalization constant, $ \epsilon_{abc} $ is the anti-symmetric Levi-Civita tensor, $a, b, c$ are color indices, $q_{1(2)}$ denotes the light quark ($ u, d $ or $ s $), $ Q $ is the bottom ($ b $) or charm ($ c $) quark and  $C$ is the charge conjugation operator. The normalization constant $A$ and the $q_{1(2)}$ quark for the considered baryons are given in Table \ref{tab:ValueofA}.

\begin{table}[thb]
	
	\renewcommand{\arraystretch}{1.3}
	\addtolength{\arraycolsep}{-0.5pt}
	\small
	$$
	\begin{array}{|l|c|c|c|c|c|c|}
	\hline \hline
	& \Sigma_{b(c)}^{*+(++)} & \Sigma_{b(c)}^{*0(+)} & \Sigma_{b(c)}^{*-(0)}  
	& \Xi_{b(c)}^{*0(+)}    & \Xi_{b(c)}^{*-(0)} 
	& \Omega_{b(c)}^{*-(0)}          \\  \hline
	A   & \sqrt{1/3} & \sqrt{2/3} & \sqrt{1/3}
	& \sqrt{2/3} & \sqrt{2/3} & \sqrt{1/3} \\  \hline
	q_1 & u & u & d & u & d & s \\  \hline
	q_2 & u & d & d & s & s & s  \\
	
	\hline \hline
\end{array}
$$
\caption{The light quark flavors for the single heavy baryons with
	spin-$ 3/2 $ and the value of normalization constant A.}
\label{tab:ValueofA}
\renewcommand{\arraystretch}{1}
\addtolength{\arraycolsep}{-1.0pt}
\end{table}
By inserting the explicit form of the interpolating current into the correlator and contracting all heavy and light quark fields via Wick's theorem, we get the corelation function in the case of $ q_{1}\neq q_{2}$ in terms of the thermal light(heavy) quark propagators, $ S_{q(Q)} $, as

\begin{widetext}
\begin{eqnarray}\label{q1difq2}
\Pi_{\mu\nu}^{OPE}(q,T) &=& -\frac{2i}{3}\epsilon_{abc}\epsilon_{a'b'c'} \int d^4 x e^{iq\cdot x}  \left\lbrace S^{cc'}_{Q} Tr [S^{ba'}_{q_{2}}\gamma_{\nu}\widetilde{S}^{ab'}_{q_{1}}\gamma_{\mu}]+S^{cc'}_{q_{1}} Tr [S^{ba'}_{Q}\gamma_{\nu}\widetilde{S}^{ab'}_{q_{2}}\gamma_{\mu}]  \right. \nonumber\\
&+&S^{cc'}_{q_{2}} Tr [S^{ba'}_{q_{1}}\gamma_{\nu}\widetilde{S}^{ab'}_{Q}\gamma_{\mu}]+S^{ca'}_{Q}\gamma_{\nu}\widetilde{S}^{bb'}_{q_{2}}\gamma_{\mu}S^{ac'}_{q_{1}}+S^{cb'}_{Q}\gamma_{\nu}\widetilde{S}^{aa'}_{q_{1}}\gamma_{\mu}S^{bc'}_{q_{2}}\nonumber\\
&+&S^{cb'}_{q_{1}}\gamma_{\nu}\widetilde{S}^{aa'}_{q_{2}}\gamma_{\mu}S^{bc'}_{Q}+S^{ca'}_{q_{1}}\gamma_{\nu}\widetilde{S}^{bb'}_{Q}\gamma_{\mu}S^{ac'}_{q_{2}}+S^{ca'}_{q_{2}}\gamma_{\nu}\widetilde{S}^{bb'}_{q_{1}}\gamma_{\mu}S^{ac'}_{Q}+S^{cb'}_{q_{2}}\gamma_{\nu}\widetilde{S}^{aa'}_{Q}\gamma_{\mu}S^{bc'}_{q_{1}}  \left. \right\rbrace  .
\end{eqnarray}
 Some extra contractions arise because of the identical particles in the case of $ q_{1}=q_{2}=q $, and the correlator is obtained as
	\begin{eqnarray}\label{q1equalq2}
	\Pi_{\mu\nu}^{OPE}(q,T) &=& \frac{i}{3}\epsilon_{abc}\epsilon_{a'b'c'} \int d^4 x e^{iq\cdot x}  \left\lbrace 2S^{cc'}_{Q} Tr [S^{bb'}_{q}\gamma_{\nu}\widetilde{S}^{aa'}_{q}\gamma_{\mu}]+2S^{cc'}_{q} Tr [S^{bb'}_{Q}\gamma_{\nu}\widetilde{S}^{aa'}_{q}\gamma_{\mu}]  \right. \nonumber\\
	&+&2S^{cc'}_{q} Tr [S^{bb'}_{q}\gamma_{\nu}\widetilde{S}^{aa'}_{Q}\gamma_{\mu}]+4S^{ca'}_{Q}\gamma_{\nu}\widetilde{S}^{ab'}_{q}\gamma_{\mu}S^{bc'}_{q}+4S^{ca'}_{q}\gamma_{\nu}\widetilde{S}^{ab'}_{q}\gamma_{\mu}S^{bc'}_{Q}\nonumber\\
	&+&4S^{ca'}_{q}\gamma_{\nu}\widetilde{S}^{ab'}_{Q}\gamma_{\mu}S^{bc'}_{q}  \left. \right\rbrace   ,
	\end{eqnarray}	
\end{widetext}
where $\widetilde{S}^{ij}_{q(Q)}=CS^{ijT}_{q(Q)}C$. 
To go further in the calculations, the thermal light quark propagator in coordinate space is selected as (see also \cite{Azizi1, Azizi2})
\begin{eqnarray}\label{lightquarkpropagator}
S_{q}^{ij}(x)&=& i\frac{\!\not\!{x}}{ 2\pi^2 x^4}\delta_{ij}-\frac{m_q}{4\pi^2 x^2}\delta_{ij}-\frac{\langle\bar{q}q\rangle_{T}}{12}\delta_{ij} \notag \\
&-&\frac{ x^{2}}{192} m_{0}^{2}\langle
\bar{q}q\rangle_{T}\Big[1-i\frac{m_q}{6}\!\not\!{x}\Big]\delta_{ij}
\nonumber\\
&+&\frac{i}{3}\Big[\!\not\!{x}\Big(\frac{m_q}{16}\langle
\bar{q}q\rangle_{T}-\frac{1}{12}\langle u^{\mu}\Theta_{\mu\nu}^{f}u^{\nu}\rangle\Big)\nonumber\\
&+&\frac{1}{3}\Big(u\cdot x\!\not\!{u}\langle
u^{\mu}\Theta_{\mu\nu}^{f}u^{\nu}\rangle\Big)\Big]\delta_{ij}
\nonumber\\
&-&\frac{ig_s \lambda_{A}^{ij}}{32\pi^{2} x^{2}}
G_{\mu\nu}^{A}\Big(\!\not\!{x}\sigma^{\mu\nu}+\sigma^{\mu\nu}
\!\not\!{x}\Big)\nonumber\\
&-&i\frac{x^2 \!\not\!{x} g_{s}^{2} \langle
	\bar{q}q\rangle_{T}^{2}}{7776}\delta_{ij}-\frac{x^4 \langle
	\bar{q}q\rangle_{T} \langle
	g_{s}^{2} G^2\rangle_{T}}{27648}+...~,\nonumber\\
\end{eqnarray}
which includes the thermal quark and gluon condensates ($\langle
	\bar{q}q\rangle_{T}  $ and $  \langle
	g_{s}^{2} G^2\rangle_{T} $), gluon fields in thermal bath, mixed condensate ($ m_{0}^{2}\langle
	\bar{q}q\rangle_{T}=\langle \bar{q}g_{s}\sigma Gq\rangle  $) as well as new operators containing the energy momentum tensor, $ \Theta_{\mu\nu} $. For the heavy quark, the following  propagator including the thermal gluon condensate and gluon fields in hot medium is used \cite{Prop_C}:
\begin{eqnarray}\label{heavypropagator}
S_{Q}^{ij}(x)&=&i \int\frac{d^4k e^{-ik \cdot x}}{(2\pi)^4} 
\left( \frac{\!\not\!{k}+m_Q}{k^2-m_Q^2}\delta_{ij}\right.\nonumber\\
&-&\frac{g_{s}G^{\alpha\beta}_{i j}}{4} \frac{\sigma^{\alpha\beta}(\!\not\!{k}+m_Q)+(\!\not\!{k}+m_Q)\sigma^{\alpha\beta} }{(k^2-m_Q^2)^{2}} 
\nonumber\\
&+&\frac{ m_{Q}}{12}\frac{k^{2}+m_{Q}\!\not\!{k}}{(k^{2}-m_{Q}^{2})^{4}} \langle g_{s}^{2} G^2 \rangle_{T}\delta_{ij}+\cdots  \Bigg) . 
\end{eqnarray}  
In Eqs. (\ref{lightquarkpropagator}) and (\ref{heavypropagator}), $m_{q(Q)}$ denotes the light(heavy) quark mass.

The thermal quark condensates, $\langle\bar{q}q\rangle_{T}$ (for $ q=u, d $) and $\langle\bar{s}s\rangle_{T}$ are parameterized in terms of the vacuum condensates, $\langle0|\bar{q}q|0\rangle$ and $\langle0|\bar{s}s|0\rangle$. For these quantities, we use the following parametrizations in terms of temperature, which are based on the lattice QCD predictions \cite{Gubler}. Note that in this study  the temperature dependence of these quantities are given  up to a temperature $ T=300~MeV$. However, we parameterize them up to $ T_{pc}\approx155 MeV $, which is considered as  the pseudo-critical temperature for the crossover phase transition at zero chemical potential. We get,
\begin{eqnarray}\label{qbarq}
\frac{\langle\bar{q}q\rangle_{T}}{\langle0|\bar{q}q|0\rangle}&=&(A_{1}e^{\frac{T}{0.025[GeV]}}+1.015),
\end{eqnarray}
and
\begin{eqnarray}\label{qbarq}
\frac{\langle\bar{s}s\rangle_{T}}{\langle0|\bar{s}s|0\rangle}&=&(A_{2}e^{\frac{T}{0.019[GeV]}}+1.002),
\end{eqnarray}
where $A_{1}=-6.534\times10^{-4}  $ and $ A_{2}=-2.169\times10^{-5}$. As we previously mentioned,   because of the choice of the thermal rest frame in Wilson expansion, the Lorentz invariance is broken. To restore that the four-velocity vector of the medium $u^{\mu}=(1,0,0,0)$ is introduced, which implies $u^2=1$ and $ q\cdot u=q_{0} $. In the rest frame of heat bath,   $  \langle u^\mu \Theta^{f,g}_{\mu\nu} u^\nu \rangle = \langle u \Theta^{f,g} u \rangle = \langle    \Theta^{f,g} _{00} \rangle = \langle  \Theta^{f,g}  \rangle $, as well.   In thermal version, as also mentioned above, new operators representing the fermionic and gluonic parts of the energy-momentum tensor arises in OPE. The fermionic part   
$\Theta^{f}_{\mu\nu}$ appears explicitly in the light-quark propagator, while the gluonic  part of the energy-momentum tensor $\Theta^{g}_{\lambda \sigma}$ appears in the expansion of the  trace of two-gluon field strength tensor in heat bath  \cite{Mallik}:
\begin{eqnarray}\label{TrGG} 
\langle Tr^c G_{\alpha \beta} G_{\mu \nu}\rangle &=& \frac{1}{24} (g_{\alpha \mu} g_{\beta \nu} -g_{\alpha
	\nu} g_{\beta \mu})\langle G^2\rangle_{T} \nonumber \\
&+&\frac{1}{6}\Big[g_{\alpha \mu}g_{\beta \nu} -g_{\alpha \nu} g_{\beta \mu} -2(u_{\alpha} u_{\mu}g_{\beta \nu} \nonumber \\
&-&u_{\alpha} u_{\nu} g_{\beta \mu} -u_{\beta} u_{\mu}
g_{\alpha \nu} +u_{\beta} u_{\nu} g_{\alpha \mu})\Big]\nonumber \\
&\times&\langle u^{\lambda} {\Theta}^g _{\lambda \sigma} u^{\sigma}\rangle.
\end{eqnarray}
The temperature dependent gluon condensate $ \langle G^2\rangle_{T} $ is parameterized in terms of the vacuum gluon condensate $\langle 0|G^{2}|0\rangle$  \cite{Gubler} as: 
\begin{eqnarray}\label{G2TLattice}
\delta \langle \frac{\alpha_{s}}{\pi}G^{2}\rangle_{T}&=&-\frac{8}{9}[ \delta T^{\mu}_{\mu}(T)-m_{u} \delta \langle\bar{u}u\rangle_{T}\nonumber \\ &-&m_{d} \delta \langle\bar{d}d\rangle_{T}-m_{s} \delta \langle\bar{s}s\rangle_{T}], 
\end{eqnarray}
where the vacuum subtracted values of the consider quantities are used as $ \delta f(T)\equiv f(T)-f(0) $ and $ \delta T^{\mu}_{\mu}(T)=\varepsilon(T)-3p(T) $: $\varepsilon(T)$ is the energy density and $ p(T) $ is the pressure. Taking into account the recent Lattice calculations \cite{Bazavov1,Borsanyi} we get the fit function of $ \delta T^{\mu}_{\mu}(T) $ as
\begin{eqnarray}\label{epsmines3p}
\frac{\delta T^{\mu}_{\mu}(T) }{T^{4}}&=&(0.020\times e^{\frac{T}{0.034[GeV]}}+0.115).
\end{eqnarray}
For the  temperature-dependent strong coupling  \cite{Kaczmarek,Morita} we utilize 
\begin{eqnarray}\label{geks2T}
g_s^{-2}(T)=\frac{11}{8\pi^2}\ln\Big(\frac{2\pi
	T}{\Lambda_{\overline{MS}}}\Big)+\frac{51}{88\pi^2}\ln\Big[2\ln\Big(\frac{2\pi
	T}{\Lambda_{\overline{MS}}}\Big)\Big],
\end{eqnarray}
where, $\Lambda_{\overline{MS}}\simeq T_{pc}/1.14$.

Alike to the physical part, the correlation function on the OPE side is  expanded in terms of the Lorentz structures as
\begin{eqnarray}
\Pi_{\mu\nu}^{OPE}(q,T)&=&\Gamma_{1}^{OPE}\!\not\!{q}g_{\mu\nu}+\Gamma_{2}^{OPE}g_{\mu\nu}\notag \\
&+&\mbox{other~structures},
\end{eqnarray}
where $\Gamma^{OPE}_{1(2)}$ is the coefficient of the selected Lorentz structure. These functions can be expressed by the help of following dispersion integral:
\begin{equation}
\label{disp_0} \Gamma^{OPE}_{1(2)} =\int_{s_{min}}^\infty ds \dfrac{\rho^{{ OPE}}_{1(2)} (s,T)}{s-q^2}+\Gamma^{nonpert}_{1(2)},
\end{equation}
where $ s_{min}=(m_{q_{1}}+m_{q_{2}}+m_{Q})^{2} $, $\rho^{ OPE}_{1(2)}(s,T)$ is the spectral density obtained via the imaginary part of the perturbative correlation function ($ pert $ in the following equation stands for the perturbative contributions)
\begin{equation}
\rho^{ OPE}_{1(2)}(s,T)=\frac{1}{\pi}\mathrm{Im}[\Gamma^{OPE, pert}_{1(2)}],
\end{equation}
and $ \Gamma^{nonpert}_{1(2)} $ represents the contributions coming from all the nonperturbative effects. In this step, our main aim is to calculate the spectral densities, corresponding to the perturbative effects in the present study,  as well as the  nonperturbative contributions to the QCD side. To this end,  the explicit forms of the heavy and light quark propagators are inserted into Eqs. (\ref{q1difq2}) and (\ref {q1equalq2}). The next step is to perform  the  standard but lengthy calculations:  These calculations contain Fourier integrals appearing in different forms, Borel transformation as well as continuum subtraction.  By matching the coefficients of the selected structures from both the physical and OPE sides of the correlation function, we find the desired sum rules:
 \begin{eqnarray}\label{residuesumrule}
 \lambda_{B_{SH}}^{2}(T)e^{-m_{B_{SH}}^2(T)/M^2}=\hat{B}\Gamma_{1}^{OPE},
 \end{eqnarray}
 and
 \begin{eqnarray}\label{residuesumrule2}
 \lambda_{B_{SH}}^{2}(T)m_{B_{SH}}(T)e^{-m_{B_{SH}}^2(T)/M^2}=\hat{B}\Gamma_{2}^{OPE},
 \end{eqnarray}
 where the functions $\hat{B} \Gamma_{1(2)}^{OPE} $ denote the $ \Gamma_{1(2)}^{OPE} $ in Borel scheme and are  given as
\begin{eqnarray}\label{Gammafunc}
\hat{B}\Gamma_{1(2)}^{OPE}=\int_{s_{min}}^{s_{0}(T)} ds \rho_{1(2)}^{{ OPE}} (s,T)e^{-s/M^2}+\hat{B}\Gamma_{1(2)}^{nonpert},\nonumber\\
\end{eqnarray}
with $ s_{0}(T) $ being the temperature-dependent continuum threshold. We will use the above  sum rules to extract the values of the mass and residue of the baryons under consideation as well as their thermal behavior in next section.

As examples, we would like to present the explicit forms of the $ \rho_{1}^{{ OPE}} (s,T) $ and $ \hat{B}\Gamma_{1}^{nonpert} $ for the $ \Sigma_{b}^{*} $  baryon. They are obtained as
\begin{widetext}
\begin{eqnarray}\label{RhoTotal}
\rho_{1}^{OPE}(s,T) &=&\frac{-1}{96 \pi ^4 \beta} \int_{0}^1 dz \Bigg\lbrace z \left(m_{b}^2+s \beta\right) \left[z \Big(3 m_{b}^2 (z+1)-12 m_{b} m_{u}+s \beta (7 z+3)\right) 
\nonumber \\
&-& 12 m_{d} (m_{b} z-2
m_{u} \beta) \Big] \Bigg\rbrace \Theta[L(s,z)],
\end{eqnarray}
\end{widetext}
and
\begin{widetext}
\begin{eqnarray}
&&\hat{B}\Gamma_{1}^{nonpert}=\frac{-1}{1152 \pi^4}\int_{s_{min}}^{s_{0}(T)} ds\int_{0}^1 dz \Bigg\lbrace -96 \pi ^2 \left\lbrace \langle\bar{d}d\rangle \Big[ 2 z (-2 m_{b}+m_{d}+2 m_{u})-3 m_{d} z^2+m_{d}-4 m_{u}\Big] \right. 
\nonumber \\
&+&\left.  \langle\bar{u}u \rangle\Big[ -4
m_{b} z+4 \beta m_{d} +m_{u} (2-3 z) z+m_{u}\Big] \right\rbrace+ g_{s}^{2} \Big(\langle G^{2}\rangle\Big[ (43-6 z) z+2\Big] 
\nonumber \\
&+& 2 \langle u \Theta^{g} u \rangle [z (21 z+23)+15]\Big)+256
\pi ^2 \beta \langle u \Theta^{f} u \rangle (5 z-1) \Bigg\rbrace  \Theta[L(s,z)]
\nonumber \\
&+&\int_{0}^1 dz e^{\frac{m_{b}^2}{M^{2} \beta}} g_{s}^{2} \Bigg\lbrace \frac{-1}{1152 \pi ^4 M^{2} \beta^2} \Big( m_{b}^2 z  \langle G^{2} \rangle \left\lbrace z (2 m_{b} (m_{d}+ m_{u} )+M^{2}) -4
m_{d} m_{u}\right\rbrace \Big) 
\nonumber \\
&+&\frac{1}{13824 \pi^2 M^{6} \beta^3}\Bigg( \langle\bar{d}d \rangle \left\lbrace \langle G^{2}\rangle \left[ -16 m_{b}^4 m_{d} z+8 m_{b}^3
\beta (m_{d} m_{u}+2 M^{2} z)-4 m_{b}^2 M^{2} \beta \Big(5 m_{d} (3 z+1)-4 m_{u}\Big) \right. \right. 
\nonumber \\
&+& \left.  8 m_{b} M^{2} \beta \Big(2
m_{d} m_{u} (3 z-2)+M^{2} (z (10 z-3)-3)\Big)+119 m_{d} M^{4} \beta^3\right]+8 M^{2} \beta^2 \langle u\Theta^{g} u\rangle \Big(6
m_{b}^2 m_{d}-12 m_{b} M^{2}
\nonumber \\
&+& \left.  m_{d} \beta \left(31 M^{2}-8 q_{0}^{2}\right)\Big) \right\rbrace + \langle\bar{u}u \rangle
\left\lbrace \langle G^{2}\rangle \left[ -16 m_{b}^4 m_{u} z+8 m_{b}^3 \beta (m_{d} m_{u}+2 M^{2} z)+4 m_{b}^2 M^{2} \beta \Big( 4
m_{d}-5 (3 m_{u} z+m_{u})\Big) \right. \right.  
\nonumber \\
&+& \left. 8 m_{b} M^{2} \beta \Big(2 m_{d} m_{u} (3 z-2)+M^{2} (z (10 z-3)-3)\Big) +119 m_{u}
M^{4} \beta^3\right] +8 M^{2} \beta^2 \langle u\Theta^{g} u\rangle \Big(6 m_{b}^2 m_{u}-12 m_{b} M^{2} 
\nonumber \\
&+& m_{u} \beta \left(31
M^{2}-8 q_{0}^{2}\right)\Big\rbrace \Bigg) +\frac{1}{663552 \pi ^4 M^{6} \beta^3}\Bigg(m_{b}^2 \langle G^{2}\rangle^2 g_{s}^{2} \Big\lbrace 32 m_{b}^2 z+M^{2} \beta (187
z+16)\Big\rbrace 
\nonumber \\
&+& 4 \langle G^{2}\rangle \Big\lbrace 64 \pi^2 \langle u\Theta^{f}u\rangle \Big[8 m_{b}^4 z-2 m_{b}^3 \beta (m_{d}+m_{u}) -2 m_{b}^2 \beta (M^{2} (z-5)-32 q_{0}^{2} z)  
\nonumber \\
&-& 4 m_{b} M^{2} \beta (3 z-2) (m_{d}+m_{u})-55 M^{4} \beta^3\Big] -m_{b}^2
g_{s}^{2} \langle u \Theta^{g} u \rangle \left(16 m_{b}^2 z \right. 
\nonumber \\
&+& \left. M^{2} \beta (85 z-16)\right)\Big\rbrace -3072 \pi ^2 M^{2} \beta^2 \langle
u \Theta^{f}u\rangle
\langle u\Theta^{g} u \rangle \Big(2 m_{b}^2+\beta \left(5 M^{2}+8 q_{0}^{2}\right)\Big)\Bigg)\Bigg\rbrace \Theta[L(s_{0},z)] 
\nonumber \\
&+&\frac{e^{-\frac{m_{b}^2}{M^{2}}}}{\pi ^2 }\Bigg\lbrace \frac{m_{0}^{2}}{72 M^{2}} \Bigg( \langle\bar{d}d\rangle [2 m_{b} m_{d} m_{u}+M^{2} (m_{d}-6
m_{u})]+\langle\bar{u}u \rangle [2 m_{b} m_{d} m_{u}+M^{2} (m_{u}-6 m_{d})]\Bigg) \nonumber \\
&-& \frac{1}{972 M^{4}} \Bigg( \langle\bar{u}u \rangle \Big[ 27 \pi ^2 \langle\bar{d}d \rangle \Big( 3 m_{b}^2 m_{d} m_{u}-8 m_{b}
M^{2} (m_{d}+m_{u})+2 M^{2} (3 m_{d} m_{u}+8 M^{2})\Big) 
\nonumber \\
&+&4 M^{2} g^{2}_{s} \langle\bar{u}u \rangle \Big(m_{b}
(m_{d}+m_{u})+M^{2}\Big)\Big]\Bigg)+\frac{1}{6912 M^{2}}  \Bigg(\langle\bar{d}d\rangle \Big[ 13 m_{d} M^{2} \langle G^{2}\rangle g^{2}_{s}+52 m_{d} M^{2}
g^{2}_{s} \langle u \Theta^{g}u \rangle 
\nonumber \\
&+&512 \pi ^2 \langle u \Theta^{f} u \rangle \Big(3 m_{b}^2 m_{d}-4 m_{b} M^{2}-4 m_{d} \left(M^{2}+2
q_{0}^{2}\right)\Big)\Big] +\langle\bar{u}u\rangle \Big[ \langle G^{2}\rangle g^{2}_{s} \Big(16 m_{b}^2 (m_{d}+m_{u})
\nonumber \\
&+&\left. M^{2} (32
m_{b}+16 m_{d}+35 m_{u})\Big)+76 m_{u} M^{2} g^{2}_{s} \langle u \Theta^{g} u \rangle+512 \pi ^2 \langle u\Theta^{f} u \rangle \Big(3 m_{b}^2
m_{u}-4 m_{b} M^{2}-12 m_{u} \left(M^{2}+2 q_{0}^{2}\right)\Big)\right] \Bigg)
\nonumber \\
&+&\frac{1}{162 M^{8}} \Bigg(3 \pi ^2 m_{0}^{2} \langle\bar{d}d \rangle \langle\bar{u}u \rangle \Big[3 m_{b}^4 m_{d} m_{u}-5
m_{b}^3 M^{2} (m_{d}+m_{u})+3 m_{b}^2 M^{2} (4 M^{2}-m_{d} m_{u})+3 M^{4} (4 M^{2}-m_{d}
m_{u})\Big]
\nonumber \\
&-&M^{4} \langle u \Theta^{f} u \rangle \Big[M^{2} \langle G^{2}\rangle g_{s}^{2}+4 M^{2} g_{s}^{2} \langle u\Theta^{g} u \rangle+16 \pi ^2
\langle u \Theta^{f} u \rangle \left(3 m_{b}^2+8 M^{2}+16 q_{0}^{2}\right)\Big]\Bigg) \Bigg\rbrace,
\end{eqnarray}
\end{widetext}
where $\Theta  $ stands for the unit-step function,  $L(s,z)=s~z(1-z)-m_{b}^2~z  $ and $ \beta=z-1 $.

\section{Numerical results}

In this section, we analyze the obtained sum rules for the masses and residues. They includes some input parameters such as the heavy and light quark masses, $ m_0^2 $, quark and gluon condensates in vacuum and energy of the quasi-particle in medium, $ q_0 $. Their numerical values are presented in Table \ref{tab:Param}.

\begin{table}[ht!] 
	\centering
	\begin{tabular}{ |c|c|}
		\hline \hline
		Parameter  &  Numeric ~Value   \\ \hline
		$q_0^{\Sigma^{*}_{b}} $; $ q_0^{\Sigma^{*}_{c}} $   &  $ (5832.1 \pm 1.9) $ $MeV$; $ (2518.48 \pm 0.20)  $ $MeV$    \\   
		$ q_0^{\Xi^{*}_{b}} $; $q_0^{\Xi^{*}_{c}} $   &  $ (5949  \pm 1.9) $ $MeV$ ; $ (2646.32  \pm 0.31) $ $MeV$      \\  
		$ q_0^{\Omega^{*}_{b}} $; $ q_0^{\Omega^{*}_{c}} $   &  $( 6.08 \pm 0.40) $ $GeV$;  $ (2765.9\pm 2.0) $  $MeV$  \\ 
		$ m_{u}   $ ;  $ m_{d}   $       &  $(2.2_{-0.4}^{+0.5})$ $MeV$; $(4.7_{-0.3}^{+0.7})$ $MeV$   \\
		$  m_{s}   $          &  $(95_{-3}^{+9} )$ $MeV$      \\
		$ m_{b}   $ ; 	$ m_{c}   $         &   $(4.18_{-0.03}^{+0.04})$ $GeV$; $(1.275_{-0.035}^{+0.025})$ $GeV$   \\
		$m_{0}^{2};   $          &  $(0.8\pm0.2)$ $GeV^2$     \\
		$ \langle0|\overline{q}q|0\rangle  (q=u, d)$          &  $-(272(5)~MeV)^3$       \\
		$ \langle0|\overline{s}s|0\rangle $          &  $-(296(11)~MeV)^3$        \\
		$ {\langle}0\mid \frac{1}{\pi}\alpha_s G^2 \mid 0{\rangle}$          &  $ 0.028(3)~GeV^4$  \\
		\hline \hline
	\end{tabular}
	\caption{Input parameters used in calculations \cite{Belyaev,Dosch,Ioffe1,pdg,Gubler}.}
	\label{tab:Param}
\end{table}

In addition, we also need to have the gluonic and fermionic parts of the energy density. Based on the lattice QCD results on the thermal behavior of the energy-momentum tensor given in \cite{Bazavov1}, their parametrizations, up to the pseudo-critical point under consideration in the present study,  are obtained as
\begin{eqnarray}\label{tetaf}
\frac{\langle\Theta^{f}\rangle}{T^{4}}&=& (0.009\times e^{\frac{T}{0.0402[GeV]}}+0.024),
\end{eqnarray}
\begin{eqnarray}\label{tetag}
\frac{\langle\Theta^{g}\rangle}{T^{4}}&=& (0.091\times e^{\frac{T}{0.047[GeV]}}-0.731),
\end{eqnarray}
 which, we are going to use them in our numerical computations.  The next problem is to obtain the parametrization of $ s_{0}(T) $ as a function of temperature. This function shall reduce to the vacuum threshold, $ s_0 $,  at zero temperature. We parameterize it as 
 \begin{eqnarray}\label{continuumthreshold}
s_{0}(T)=s_{0} f(T),
\end{eqnarray}
such that at $T \rightarrow 0$ limit, $f(T) \rightarrow 1 $. Hence, we should first determine $ s_0 $ based on the standard prescriptions of the method,  afterwards we will extract the function $f(T)  $ from the calculations.

\begin{widetext}

\begin{figure}[ht]
	\begin{center}
		\includegraphics[width=12.5cm]{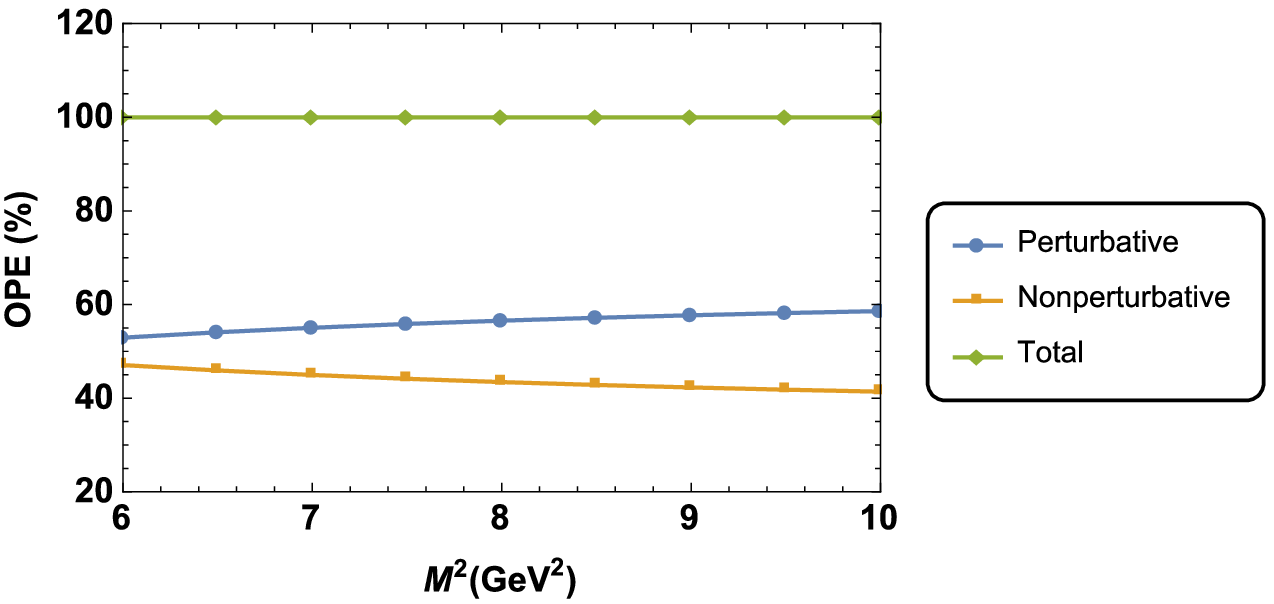}
		\includegraphics[width=12.5cm]{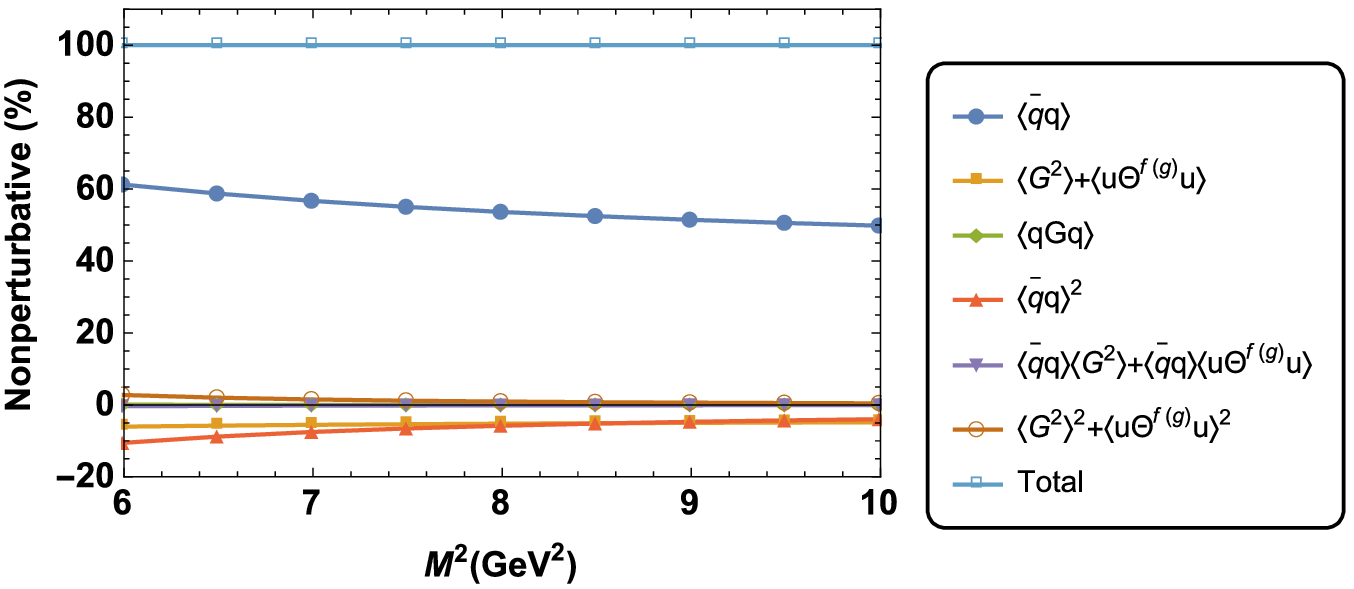}		
	\end{center}
	\caption{Up: Contributions of perturbative and nonperturbative parts to total OPE. Down: Contributions of various operators with different dimensions to nonperturbative part:  $\langle \bar{q}q \rangle $ (dimension 3), $\langle G^{2}\rangle + \langle u \Theta^{f(g)} u \rangle  $ (dimension 4), $ \langle qGq \rangle $ (dimension 5) , $ \langle \overline{q}q\rangle^{2} $ (dimension 6), $\langle \overline{q}q\rangle \langle G^{2}\rangle+ \langle \overline{q}q\rangle \langle u \Theta^{f(g)} u \rangle) $ (dimension 7), $  \langle G^{2}\rangle^{2}+ \langle u \Theta^{f(g)} u \rangle^{2} $(dimension 8). } \label{fig2}
\end{figure}

\end{widetext}

Besides the  continuum threshold in vacuum the sum rules obtained in previous section include another auxiliary parameter, Borel parameter $ M^{2} $, which should also be fixed. We need to determine the working regions of  $ s_0 $  and $ M^{2} $ such that the physical quantities under consideration show mild dependence on these parameters. The  continuum threshold $ s_{0} $ is not totally free but it is related to the energy of the first excited state in the same channel. Thanks to the experiments that have provided many new results not only on the ground states but also on the excited states of some single heavy baryons, recently \cite{pdg}.  In view of PDG, we see that the excited states generally have energies about $ 300 ~MeV $ above the ground states masses. In choosing the working window for the $ s_{0} $, we also  look after  the pole dominance and OPE convergence in our sum rules.  These considerations leads to the window: 
\begin{equation}
[m_{B_{SH}}+0.3]^{2}~GeV^2\leq s_0\leq[m_{B_{SH}}+0.5]^{2}~GeV^2. 
\end{equation}
The upper and lower limits of the Borel parameter are fixed consider the criteria of the QCD sum rule method. To find the lower limit, we apply the  criterion of the OPE convergence at the chosen window for the continuum threshold. To this end, we demand that the perturbative part exceeds the total nonperturbative contributions and the slogan of  \textit{ the higher the dimension of the nonperturbative operator the lower its contribution} is satisfied. Our calculations show that the operators having eight dimensions, the higher dimension that we include into the analyses,  constitute only one percent of the total contribution at lower value of $ M^{2} $, i.e. $ \Gamma_{1(2)}^{8,OPE}(M_{min}^{2},s_{0})/\Gamma_{1(2)}^{total, OPE}(M_{min}^{2},s_{0}) \simeq 0.01 $.  Figure \ref{fig2} shows the perturbative and nonperturbative contributions to total OPE as well as the  contributions of different nonperturbative operators with various mass dimensions, separately. This figure  depicts a nice convergence of the OPE in our calculations. As it is clear, the perturbative contribution dominates over nonperturbative contributions and it is about $ 53\%$ of the total  at $ M^{2}_{min}=6~GeV^2 $. The main contribution in nonperturbative part belongs to the quark condensate, $ \langle \overline{q}q\rangle $.

\begin{figure}[ht]
	\begin{center}
	\includegraphics[width=8cm]{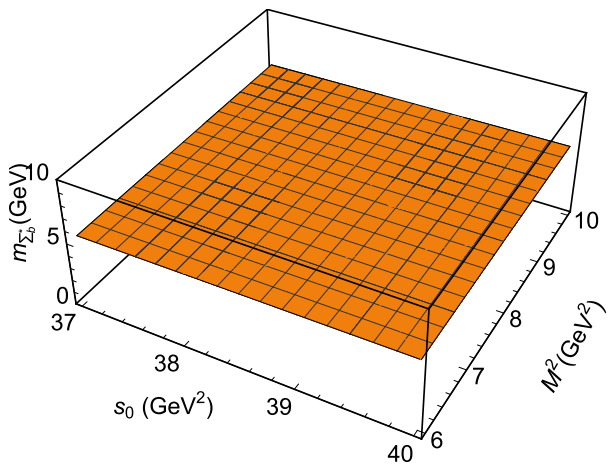}
	\end{center}
	\caption{The mass of the $\Sigma_{b}^{*}$ baryon as  functions of $M^2$ and $ s_{0} $ at $T=0$.} \label{fig3}
\end{figure}

To obtain $ M^{2}_{max} $, we utilize the condition of the  pole dominance as 
\begin{eqnarray}
PC=\frac{ \Gamma_{1(2)}^{OPE}(M^{2},s_{0})}{ \Gamma_{1(2)}^{OPE}(M^{2},\infty)}\geq\frac{1}{2}.
\end{eqnarray}
As a result, we get the working region of the Borel parameter as $ M^{2}\in[6,10] ~GeV^2$. We plot,  as an example,  a 3D graphic of the mass of $\Sigma_{b}^{*}$ baryon as  functions of $M^2$ and $ s_{0} $ at $T=0$ in Figure \ref{fig3}.   As is seen the mass shows good stability against the variations of the auxiliary parameters in the selected windows. 

\begin{widetext}

	\begin{figure}[ht]
		\begin{center}
		\subfigure[]{\includegraphics[width=8cm]{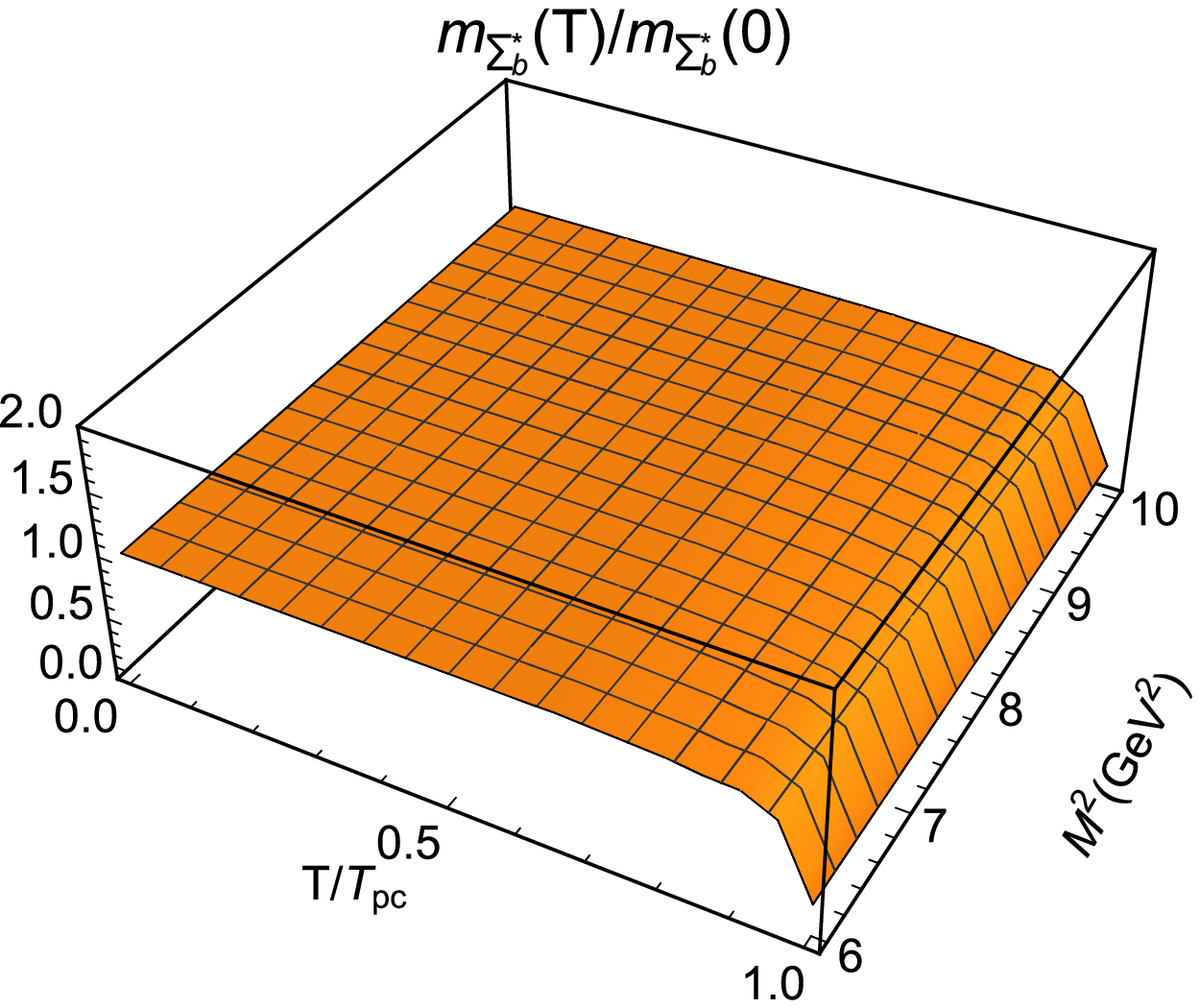}}
			\subfigure[]{\includegraphics[width=8cm]{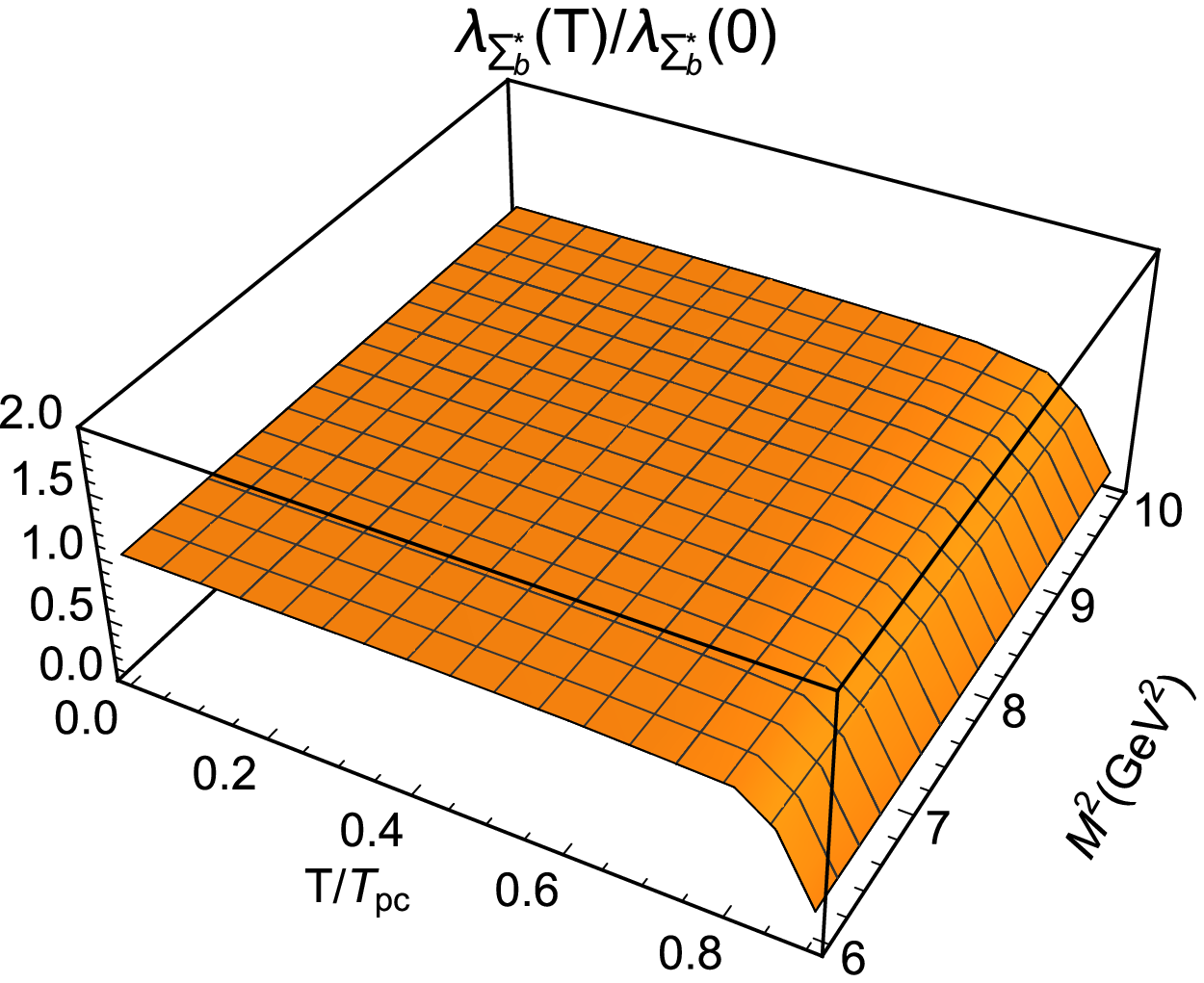}}
		\subfigure[]{\includegraphics[width=8cm]{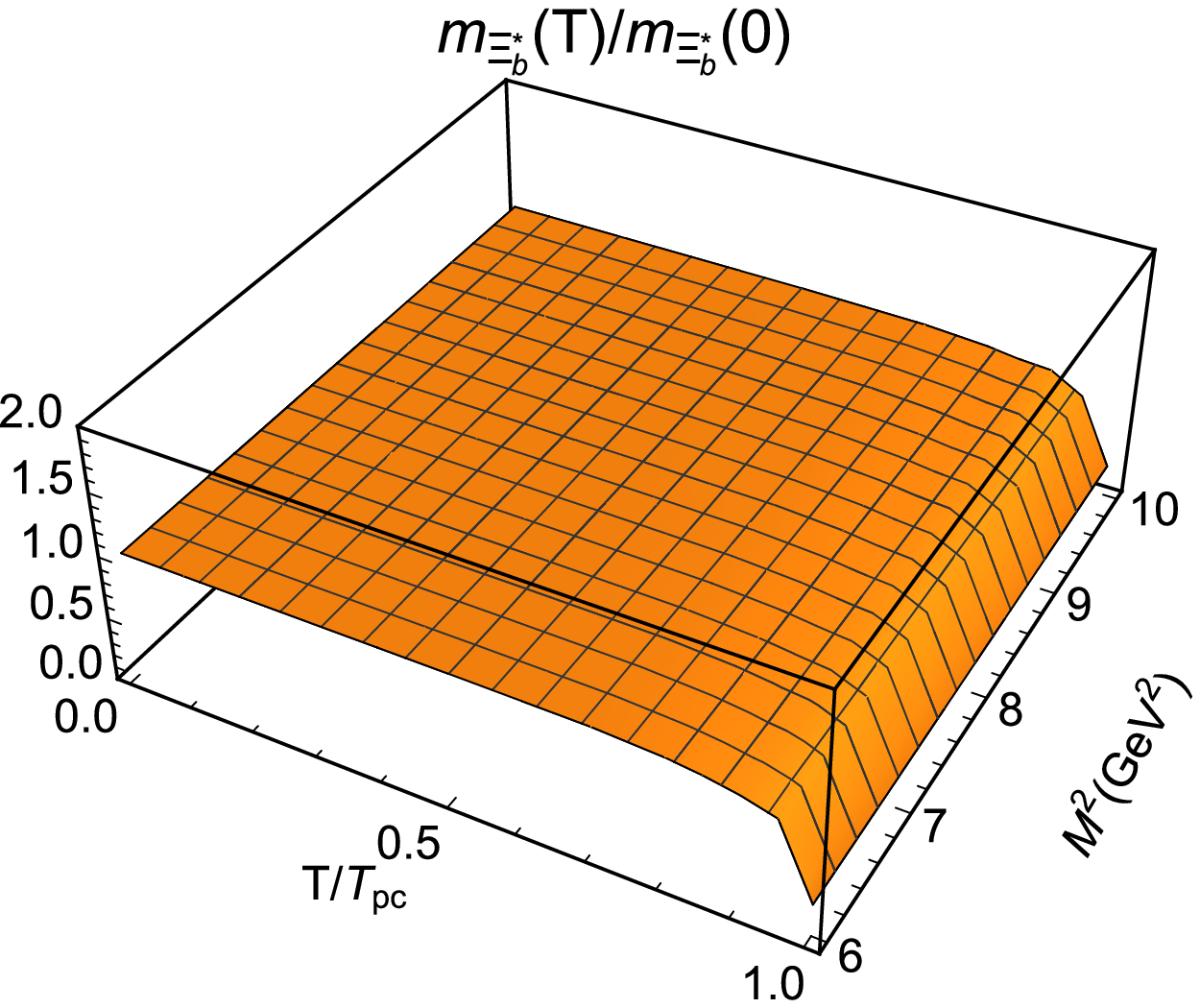}}
			\subfigure[]{\includegraphics[width=8cm]{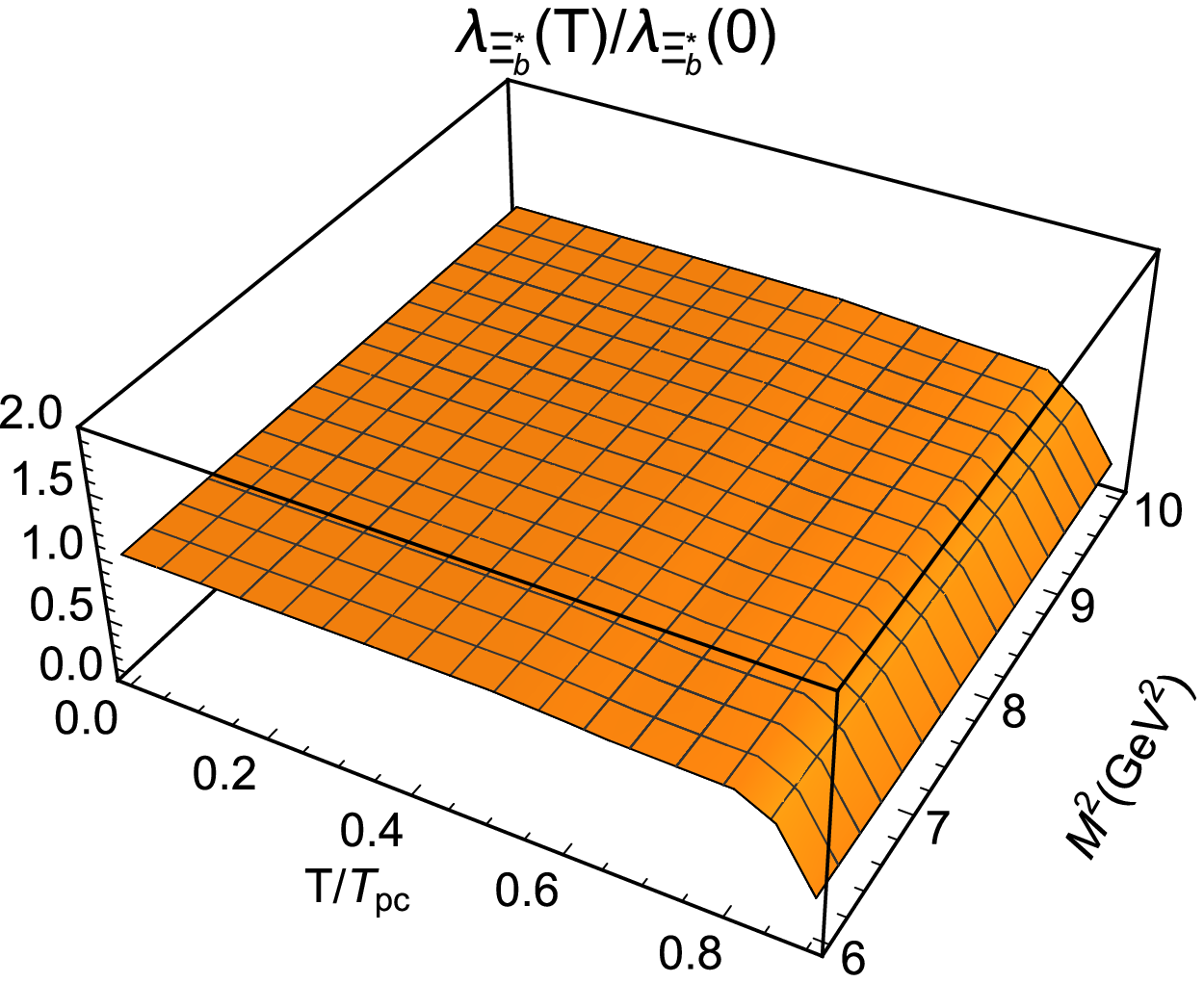}}
			\subfigure[]{\includegraphics[width=8cm]{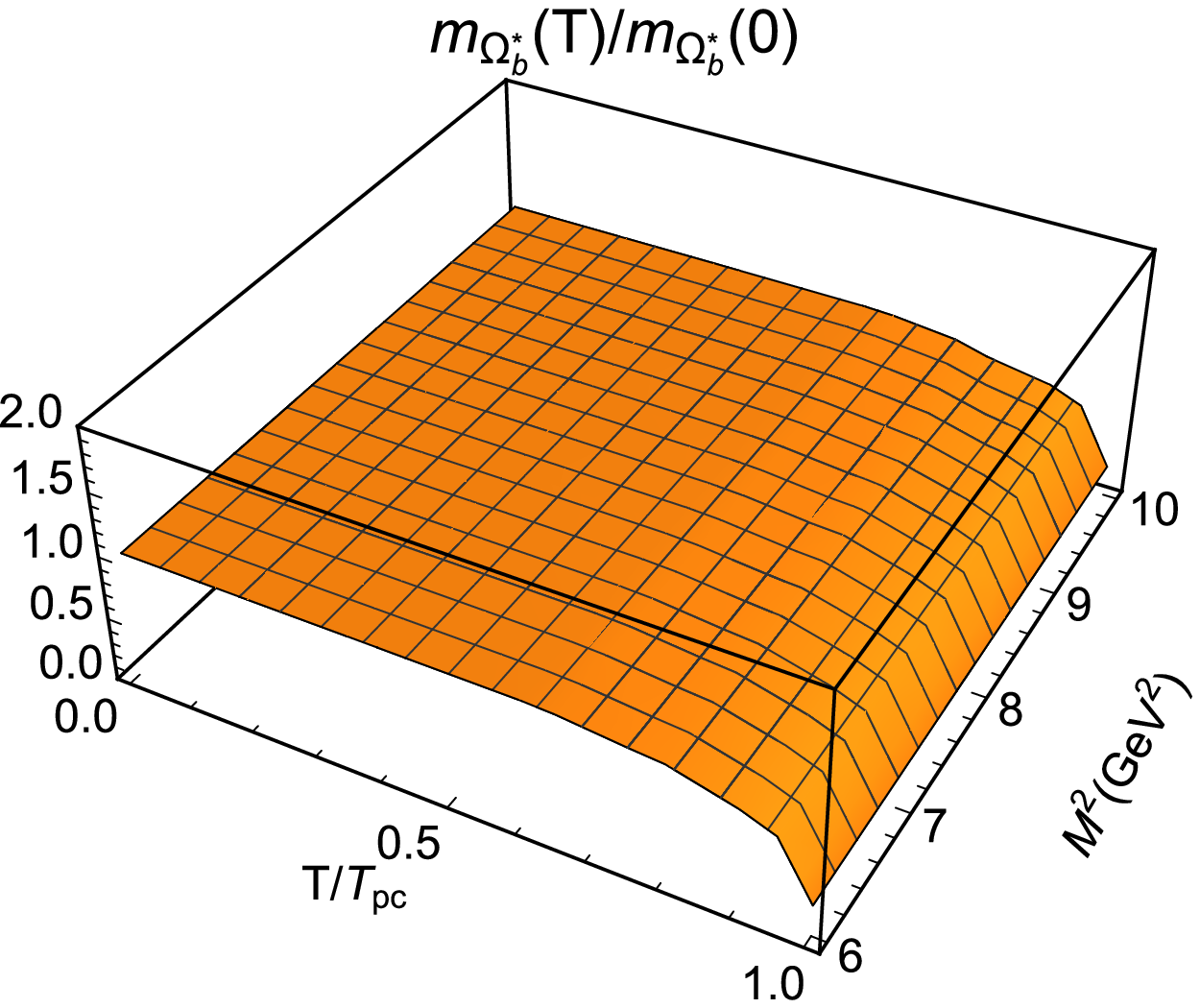}}
		\subfigure[]{\includegraphics[width=8cm]{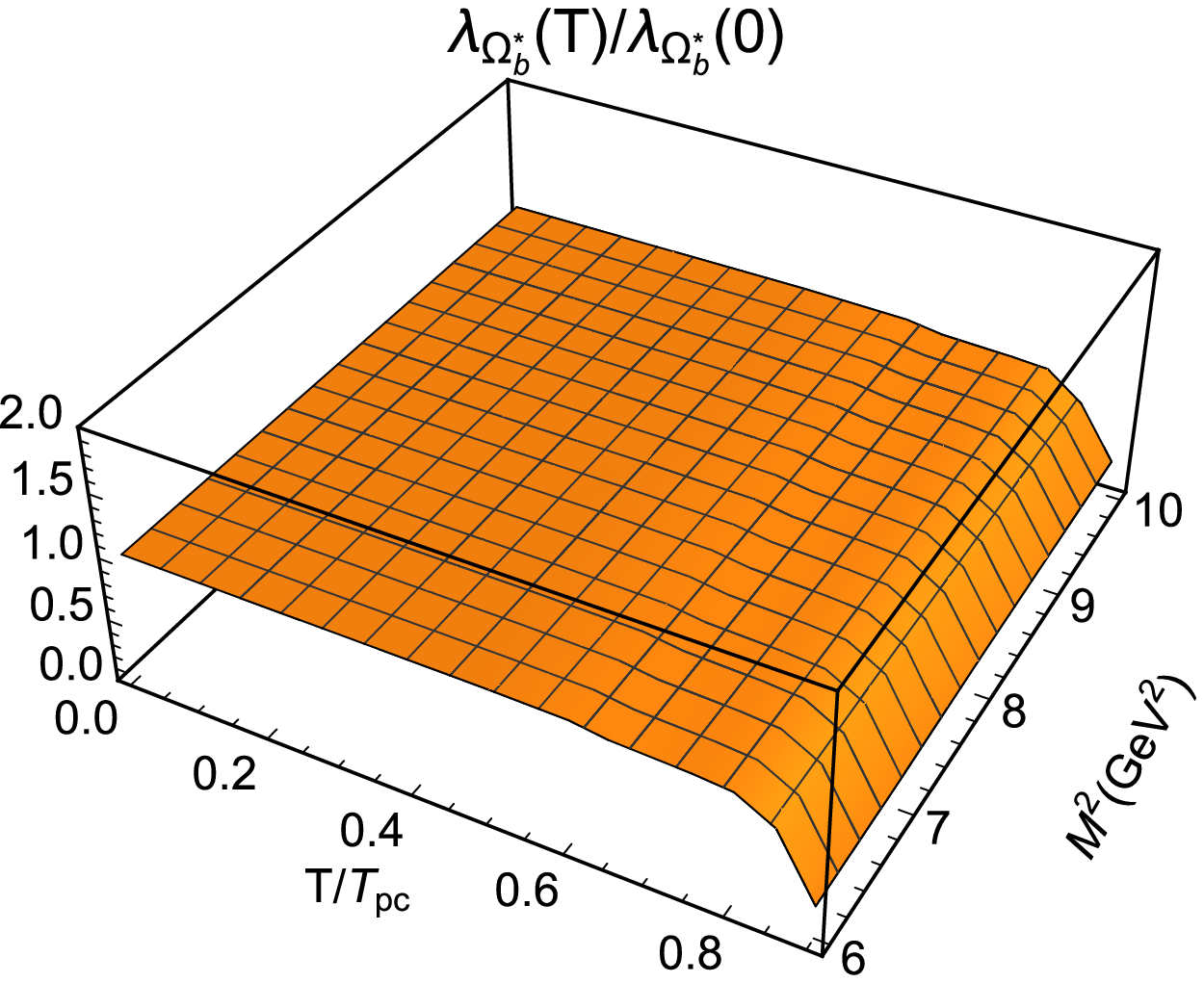}}
		\end{center}
		\caption{The mass (right) and residue (left) of the bottom baryons as  functions of $M^2$ and $T/T_{pc}$.} \label{fig4}
	\end{figure}
	
	\end{widetext}

Now, we proceed to find the function $ f(T) $ and the temperature dependent mass $ m_{B_{SH}}(T) $ and residue $ \lambda_{B_{SH}}(T) $ of the single heavy spin-3/2 baryons. To this end, we use the two sum rules in Eqs. (\ref{residuesumrule}) and (\ref{residuesumrule2}) and one extra equation obtained by applying the derivative with respect to $ \frac{d}{d(-\frac{1}{M^2})} $  to both sides of  Eq. (\ref{residuesumrule}). Simultaneous solving of the resultant three equations with the aim of obtaining the three mentioned unknowns gives the function $ f(T) $ as
\begin{eqnarray}\label{fTcontinuumthreshold}
f(T)=1-0.96\Big(\frac{T}{T_{pc}}\Big)^{9}.
\end{eqnarray}
In the following, we proceed to discuss the thermal behavior of the masses and residues under study as the main goal of the present work. 
In this context, as examples,  we plot the $ m(T)/m(0) $ and $ \lambda(T)/\lambda(0) $ for the bottom members  as  functions of $ T/T_{pc} $ and $M^2$ in Figure \ref{fig4} at average value of the vacuum continuum threshold. This figure shows that the spectroscopic parameters of the $ \Sigma_{b}^{*}, \Xi_{b}^{*}$ and $ \Omega_{b}^{*} $ baryons are stable against the changes in temperature until a certain temperature but after that, they start to decrease with increasing the temperature. Our analyses show that the charmed baryons present similar behavior, as well.  The points that the stability starts to break down for mass and  residue are  $T\cong0.14~ GeV$ and $T\cong0.13~ GeV$, respectively. After  these points the mass and residue starts to diminish. The mass and residue 
fall  substantially near to the pseudo-critical temperature. The amount of decrements at $ T_{pc} $ are  $75\%$ ( $66-71\%$) for the mass of bottom (charmed)  and  $71-80\%$ ( $42-50\%$) for the residue of bottom (charmed) baryons, respectively compared to their vacuum values. These behavior of baryons  can be interpreted as substantial melting of the heavy baryons near to the  pseudo-critical temperature. 

	At the end of this section, we would like to present our results for the masses of the single heavy spin-3/2 baryons at $ T\rightarrow 0 $ limit. This is done in table \ref{tab:1}. For comparison, we also present the existing theoretical predictions in the literature  and experimental  data in the same table.  With a quick glance in this table, we see that our predictions, within the errors, are overall consistent with other theoretical predictions made using different methods and approaches. Our predictions are also well consistent with the existing experimental data for five members within the presented uncertainties. $\Omega_{b}^{*}$ baryon is only missing member, which has not been discovered in the experiment. We hope that, our result together with other theoretical predictions will help experimental group in the course of search for this particle.
	
	\begin{widetext}
	
\begin{table}[h]
\centering
\begin{tabular}{|c||c|c|c|c|c| c|}\hline
&$m_{\Omega_{b}^{*}}$ & $m_{\Omega_{c}^{*}}$& $m_{\Sigma_{b}^{*}}$ &$m_{\Sigma_{c}^{*}}$  & $m_{\Xi_{b}^{*}}$& $m_{\Xi_{c}^{*}}$\\\cline{1-7}
\hline\hline
present work&$6.08^{+0.10}_{-0.15}$&$2.75^{+0.08}_{-0.26}$ &$5.88^{+0.11}_{-0.11}$&$2.56^{+0.08}_{-0.07}$&$5.95^{+0.12}_{-0.13}$&$2.65^{+0.08}_{-0.07}$\\\cline{1-7}
\cite{Capstick}&-&-&5.805&2.495&-&-\\\cline{1-7}
\cite{Bagan}&-&-&$ 5.4\sim6.2 $&$ 2.15\sim2.92 $&-&-\\\cline{1-7}
\cite{Savage}&-&2.768&-&2.518&-&-\\\cline{1-7}
\cite{Roncaglia}&$ 6.090\pm0.050 $&$ 2.770\pm 0.030$&$ 5.850 \pm 0.040$&$ 2.520\pm0.020 $&$ 5.980 \pm0.040$&$ 2.650 \pm0.020$\\\cline{1-7}
\cite{Jenkins}&6.083&2.760&5.840&-&5.966&-\\\cline{1-7}
\cite{Dai}&-&-&$ 5.84\pm0.09 $&$2.55\pm0.08 $&-&-\\\cline{1-7}
\cite{Wang}&-&-&$ 5.82\pm0.13 $&$2.59\pm0.20 $&-&-\\\cline{1-7}
\cite{Mathur}&6.060&2.752&5.871&2.538&5.959&2.680\\\cline{1-7}
\cite{Ebert}&6.088 &2.768 &5.834&2.518&5.963&2.654\\\cline{1-7}
\cite{Zhang,Zhang1}&$6.00\pm0.16$&$2.74\pm0.23$ &$5.81\pm0.19$&$2.56\pm0.24$&$5.94\pm0.17 $&$2.64\pm0.22$\\\cline{1-7}
\cite{Wang2}&$ 6.06\pm0.13 $&$ 2.76\pm0.10 $&-&-&-&-\\\cline{1-7}
\cite{Valcarce}&$ 6.079 $&$ 2.767 $&5.829&2.502&5.961&2.642\\\cline{1-7}
\cite{Liu}&$6.063^{+0.083}_{-0.082}$&$2.790^{+0.109}_{-0.105}$
&$5.835^{+0.082}_{-0.077}$&$2.534^{+0.096}_{-0.081}$&$5.929^{+0.083}_{-0.079}$&$2.634^{+0.102}_{-0.094}$\\\cline{1-7}
\cite{Gerasyuta}&-&-&5.829&-&-&-\\\cline{1-7}
\cite{Karliner}&6.082&-&-&-&$ 5.959\pm0.004 $&-\\\cline{1-7}
\cite{Aliev}&$6.08\pm0.40$&$2.72\pm0.20$ &$5.85\pm0.35$&$2.51\pm0.15 $&$5.97\pm0.40 $&$2.66\pm0.18 $\\\cline{1-7}
\cite{Lewis}&$6.044\pm0.018$&- &$5.842\pm0.026$&-&$5.950\pm0.021 $&-\\\cline{1-7}
\cite{WangZG}&$ 6.17\pm0.15 $&$ 2.79\pm0.19 $&$5.85\pm0.20 $&$2.48\pm0.25 $&$6.02\pm0.17 $&$2.65\pm0.20 $\\\cline{1-7}
\cite{Ebert1}&$ 6.088 $&$ 2.768 $&$5.834 $&$2.519 $&$5.963 $&$2.649 $\\\cline{1-7}
\cite{Kim}&$ 6.073 $&-&$5.834 $&-&$5.954 $&-\\\cline{1-7}
Exp\cite{Nakamura}&-&$2.7659\pm 0.0020 $&$5.83032\pm 0.00027$&$2.51848\pm 0.00020$&$5.9523\pm 0.0009$&$2.64638\pm 0.00021$\\\cline{1-7}
\end{tabular}
\vspace{0.8cm}
\caption{The vacuum mass comparison of the single heavy spin-3/2 baryons in $GeV$ with existing theoretical predictions and experimental  data (Exp\cite{Nakamura}).   
}\label{tab:1}
\end{table}

\end{widetext}

\label{sec:results}
\section{Summary and Concluding remarks}

In this study, we have performed two-point thermal QCD sum rule analyses for $ \Sigma_{Q}^{*},~ \Xi_{Q}^{*}$ and $ \Omega_{Q}^{*} $ single heavy baryons  which are the members of the spin-$ 3/2 $ sextet family. In the OPE,  operators up to dimension eight were taken into account which lead to a good OPE convergence as well as pole dominance. We included the thermal effects by two ways: We replaced the vacuum condensates by their thermal versions and considered the extra operators, appearing in the forms of the fermionic and gluonic parts of the energy momentum tensor due to the restoration of the Lorentz invariance. We fixed the auxiliary parameters entering the calculations by the standard prescriptions of the method. By simultaneous  solving of the two sum rules obtained together with an extra equation derived from one of the sum rules, we found three unknowns: Thermal continuum threshold, temperature-dependent mass and temperature-dependent residue. We discussed the thermal behavior of the mass and residue for all the bottom and charmed baryon members having the spin-3/2. We observed that the spectroscopic parameters remain unchanged up to a certain temperature, after which they start to diminish considerably near to the pseudo-critical temperature. The decrements order in the mass and residue of the considered baryons near to the pseudo-critical temperature are obtained as  $(66-75)\%$ and $(42-80)\%$, respectively, representing substantial melting of the heavy baryons near to the  pseudo-critical temperature. In the literature, there are no other studies on the thermal behavior of single heavy baryons to make a compression with our predictions. However, there are some studies on the temperature dependence of the masses of light baryons, In Refs. \cite{Azizi1, Kaya} the authors investigated the light octet and decuplet baryons using the thermal QCD sum rule, but considering a pseudo-critical temperature of $ T_{pc}=197 MeV $.  They obtained that the shifts in the masses of the considered baryons are overall about $ 80\%  $. The pole mass of the octet and decuplet baryons were also evaluated in Ref. \cite{Bedaque} via the   chiral perturbation theory. The authors observed that  a $ 20\% $ mass shift occurs around the temperature $ T\backsimeq 150 MeV $,  where the freeze-out in the relativistic heavy-ion collision is expected to be formed. Using the many-body techniques at finite temperature, all baryonic states of the octet and decuplet flavors were examined in Ref. \cite{Rincon}. They obtained that the baryon masses decrease with the temperature and there are strong dependencies on the melting (or deconfinement) temperature depending on the flavor content of the baryons. In the framework of the thermal QCD sum rule, the masses of the decuplet baryons were also investigated in Ref. \cite{Xu}. According to this  study, the masses of the decuplet baryons show very little temperature dependence  below $ T= 0.11~GeV $ and  the melting or hadron-quark phase transition occurs at a temperature $ T \geq 0.11~GeV $.  Our results indicate that this point is $ T= 0.14~GeV $ for heavy baryons, after which the masses start to decrease with the increasing of the temperature and the dependence of the masses on temperature near to the critical temperature is very strong. These information on the behavior of the masses of different baryons may help experimental groups in the analyses of the results of the  in-medium and heavy ion collision experiments, despite the statistical hadronization model claims that any thermal modification of masses is negligibly small at pseudo-critical temperature and the  in-medium mass shifts at $ T_{pc} $ would be excluded.

We extracted the values of the masses for both the bottom and charmed baryons at $ T\rightarrow 0 $ limit and compared with the predictions of other phenomenological models and experimental data. The obtained results are well consistent with existing experimental data. Our result on the mass of $\Omega_{b}^{*}$ baryon as the only undiscovered member together with other predictions may help the experimental group to hunt this particle and measure its parameters. 




%
%


\newpage
\begin{thebibliography}{99}


\bibitem{Isgur} N. Isgur and M. B. Wise, Phys. Lett. B  {\bf 232}, 113, (1989), N. Isgur and M. B. Wise, Phys. Rev. Lett. {\bf 66}, 1130, (1991).

\bibitem{Georgi} H. Georgi, Phys. Lett. B {\bf 240}, 447 (1990).

\bibitem{Nakamura}  M. Tanabashi et al. (Particle Data Group), Phys. Rev. D {\bf 98}, 030001 (2018).


\bibitem{CDF} D. Acosta et al., (CDF Collaboration), Phys. Rev. Lett. {\bf 96}, 202201 (2006). 


\bibitem{Babar} B. Aubert et al., (BABAR Collaboration), Phys. Rev. Lett. {\bf 97}, 232001 (2006).

\bibitem{CDF2} T. Aaltonen et. al., (CDF Collaboration), Phys. Rev. Lett. {\bf 99}, 202001 (2007).

\bibitem{D0} V. Abazov et. al., (D0 Collaboration), Phys. Rev. Lett. {\bf 99}, 052001 (2007).

\bibitem{CDF3} T. Aaltonen et. al., (CDF Collaboration), Phys. Rev. Lett. {\bf 99}, 052002 (2007).

\bibitem{Chistov} R. Chistov et al., (Belle Collaboration), Phys. Rev. Lett. {\bf 97}, 162001 (2006).

\bibitem{Aubert} B. Aubert et al., (BABAR Collaboration), Phys. Rev. D {\bf 77}, 012002 (2008).

\bibitem{Lesiak} T. Lesiak et al., (Belle Collaboration), Phys. Ltt. B {\bf 665}, 9 (2008).

\bibitem{CMS} S. Chatrchyan et al., (CMS Collaboration), Phys.Rev.Lett. {\bf 108}, 252002 (2012).

\bibitem{LHCb} R. Aij et al., (LHCb Collaboration), JHEP {\bf 1605}, 161 (2016).

\bibitem{Shuryak} E. V. Shuryak, Nucl. Phys. B {\bf 198}, 83 (1982).

\bibitem{Capstick} S. Capstick and N. Isgur, Phys. Rev. D {\bf 34}, 2809 (1986).

\bibitem{Bagan} E. Bagan, M. Chabab, H. G. Dosch, and S. Narison, Phys. Lett. B {\bf 278}, 367 (1992).

\bibitem{Grozin} A. G. Grozin and O. I. Yakovlev, Phys. Lett. B {\bf 285}, 254 (1992).

\bibitem{Savage} M. J. Savage, Phys. Lett. B {\bf 359}, 189 (1995).

\bibitem{Roncaglia} R. Roncaglia, D. B. Lichtenberg, and E. Predazzi, Phys. Rev. D {\bf 52}, 1722 (1995); R. Roncaglia, A. Dzierba, D. B. Lichtenberg, and E. Predazzi, Phys. Rev. D {\bf 51}, 1248 (1995).
\bibitem{Jenkins} E. Jenkins, Phys. Rev. D54, 4515 (1996).

\bibitem{Dai} Y. B. Dai, C. S. Huang, C. Liu and C. D. Lu, Phys. Lett. B {\bf 371}, 99 (1996).

\bibitem{Groote} S. Groote, J. G. K{\"o}rner and O. I. Yakovlev, Phys. Rev. D {\bf 55}, 3016 (1997).

\bibitem{Wang} D. W. Wang, M. Q. Huang and C. Z. Li, Phys. Rev. D {\bf 65}, 094036 (2002).

\bibitem{Mathur} N. Mathur, R. Lewis, and R. M. Woloshyn, Phys. Rev. D {\bf 66}, 014502 (2002).

\bibitem{Wang1} D. W. Wang, M. Q. Huang, Phys.Rev. D {\bf 67}, 074025 (2003).

\bibitem{Ebert} D. Ebert, R. N. Faustov, and V. O. Galkin, Phys. Rev. D {\bf 72}, 034026 (2005).

\bibitem{Garcilazo} H. Garcilazo, J. Vijande and A. Valcarce, J. Phys. G {\bf 34}, 961 (2007).

\bibitem{Zhang} J. R. Zhang and M. Q. Huang, Phys. Rev. D {\bf 78}, 094015 (2008).

\bibitem{Wang2} Z. G. Wang, Eur. Phys. J. C {\bf 54}, 231 (2008).


\bibitem{Roberts} W. Roberts and M. Pervin, Int. J. Mod. Phys. A {\bf 23}, 2817 (2008).

\bibitem{Valcarce} A. Valcarce, H. Garcilazo and J. Vijande, Eur. Phys. J. A {\bf 37}, 217 (2008).

\bibitem{Liu} X. Liu, H. X. Chen, Y. R. Liu, A. Hosaka, and S. L. Zhu, Phys. Rev. D {\bf 77}, 014031 (2008).

\bibitem{Groote1} S. Groote, J.G. Korner, A.A. Pivovarov, Eur.Phys.J.C {\bf 58}, 355 (2008).

\bibitem{Zhang1} J. R. Zhang and M. Q. Huang, Phys. Rev. D {\bf 77}, 094002 (2008).

\bibitem{Gerasyuta} S. M. Gerasyuta, E. E. Matskevich, Int.J. Mod.Phys. E {\bf 18}, 1785 (2009).

\bibitem{Karliner} M. Karliner, B. Keren-Zura, H. J. Lipkin, and J. L.Rosner, Annals Phys. {\bf 324}, 2 (2009).


\bibitem{Aliev} T. M. Aliev, K. Azizi, and A. Ozpineci, Nucl. Phys. B {\bf 808},137 (2009).

\bibitem{Lewis} R. Lewis, R.M. Woloshyn, Phys. Rev. D {\bf 79}, 014502 (2009).

\bibitem{WangZG} Z. G. Wang, Eur. Phys. J. C {\bf 68}, 459 (2010).


\bibitem{Ebert1} D. Ebert, R. N. Faustov and V. O. Galkin, Phys. Rev. D {\bf 84}, 014025 (2011).

\bibitem{Kim} J.Y. Kim, H. C. Kim, G.S. Yang, Phys. Rev. D {\bf 98}, 054004 (2018).

\bibitem{Azizi} K. Azizi, N. Er, Nucl.Phys. A {\bf 970}, 422 (2018).

\bibitem{Aoki}Y. Aoki, G. Endrodi, Z. Fodor, S.D. Katz, K.K. Szabo, Nature {\bf 443}, 675-678, (2006).

\bibitem{MCheng} M. Cheng et al., Phys. Rev. D {\bf 74}, 054507 (2006).

\bibitem{Bhattacharya} T. Bhattacharya et al., Phys. Rev. Let. (PRL) {\bf 113}, 082001 (2014).

\bibitem{Bazavov2} A. Bazavov et al., Phys. Rev. D {\bf 95}, 054504 (2017).

\bibitem{Shifman} M. A. Shifman, A. I. Vainstein,  V. I. Zakharov, Nucl. Phys. B \textbf{147}, 385 (1979);
M. A. Shifman,  A. I.  Vainstein, V. I. Zakharov,  Nucl. Phys. B \textbf{147}, 448 (1979).

\bibitem{Ioffe} B. L. Ioffe, Nucl. Phys. B {\bf 188}, 317 (1981).

\bibitem{Bochkarev} A. I. Bochkarev,  M. E. Shaposhnikov, Nucl. Phys. B {\bf 268}, 220 (1986).

\bibitem{Aliev1} T. M. Aliev, M. Savci, Phys.Rev. D {\bf 90}, 116006, 11 (2014).

\bibitem{Savvidy} K. G. Savvidy, (2005) [arXiv:1005.3455 [hep-th]].

\bibitem{Aliev2} T. M. Aliev, K. Azizi and M. Savci,Phys. Rev. D {\bf 82}, 096006 (2010).


\bibitem{Aliev3} T. M. Aliev, K. Azizi and M. Savci, Phys. Lett. B {\bf 681}, 240 (2009) .

\bibitem{Lee} F. X. Lee, Phys. Rev. D \textbf{57}, 1801 (1998).

\bibitem{Azizi1} K. Azizi, G. Kaya, J. Phys. G {\bf 43}, no.5, 055002 (2016).

\bibitem{Azizi2} K. Azizi, A. T\" urkan, E. Veli Veliev, H. Sundu,  Adv. High Energy Phys.  {\bf 2015}, 794243 (2015). 
\bibitem{Prop_C}L. J. Reinders, H. Rubinstein and S. Yazaki, Phys. Rept. {\bf 127}, 1 (1985).

\bibitem{Gubler} P. Gubler, D. Satow, Prog. Part. Nucl. Phys. \textbf{106}, 1 (2019) [arXiv:1812.00385 [hep-ph]].

\bibitem{Mallik} S. Mallik, Phys. Lett. B {\bf 416}, 373 (1998).

\bibitem{Bazavov1} A. Bazavov et al., Phys. Rev., D \textbf{90}, 094503 (2014).

\bibitem{Borsanyi} S. Borsanyiet al., Phys. Lett. B \textbf{730}, 99–104 (2014).




\bibitem{Kaczmarek} O. Kaczmarek, F. Karsch, F. Zantow, P. Petreczky, Phys. Rev. D {\bf 70}, 074505 (2004).

\bibitem{Morita} K. Morita, S. H. Lee , Phys. Rev. C {\bf 77}, 064904 (2008).

\bibitem{Belyaev} V. M. Belyaev, B. L. Ioffe, Sov. Phys. JETP, {\bf 57}, 716 (1983).

%
\bibitem{Dosch} H. G. Dosch, M. Jamin and S. Narison, Phys. Lett. B {\bf 220}, 251 (1989). 

\bibitem{Ioffe1} B. L. Ioffe, Prog. Part. Nucl. Phys. {\bf 56}, 232 (2006).

\bibitem{pdg}M. Tanabashi et al., Phys. Rev. D {\bf 98}, 030001 (2018).




%
%
%
%
%

%
%


%

%
%
%

%
%

\bibitem{Kaya} K. Azizi, G. Bozkir, Eur. Phys. J.  C {\bf 76}, 521 (2016).  
\bibitem{Bedaque} P. F. Bedaque, Phys. Lett. B {\bf 387}, 1 (1996). 
\bibitem{Rincon} J. M. Torres-Rincon, B. Sintes, J. Aichelin,  Phys. Rev. C {\bf 91}, 065206 (2015). 
\bibitem{Xu} Y. Xu, Y. Liu, M. Huang, Commun. Theor. Phys. {\bf 63}, 209 (2015). 


\end{thebibliography}
\end{document}